\documentclass[journal,comsoc]{IEEEtran}

\usepackage[utf8]{inputenc}
\usepackage{framed, color}
\usepackage{url}
\usepackage{amsmath}
\usepackage{amssymb}
\usepackage{multirow}
\usepackage{listings}
\usepackage{xspace}
\usepackage{epsfig}
\usepackage{graphics}
\usepackage{subcaption}
\usepackage{microtype}
\usepackage{textcomp}
\usepackage{times}
\usepackage{caption}
\makeatletter
\newcommand\nocaption{%
    \renewcommand\p@subfigure{}
    \renewcommand\thesubfigure{\thefigure\alph{subfigure}}
}
\makeatother

\newtheorem{lemma}{Lemma}

\begin{document}

\title{Re-designing Dynamic Content Delivery in the Light of a Virtualized Infrastructure}

\author{\IEEEauthorblockN{
Giuseppe Siracusano\IEEEauthorrefmark{1},
Roberto Bifulco\IEEEauthorrefmark{2},
Martino Trevisan\IEEEauthorrefmark{3},
Tobias Jacobs\IEEEauthorrefmark{2},
Simon Kuenzer\IEEEauthorrefmark{2},
Stefano Salsano\IEEEauthorrefmark{1},
Nicola Blefari-Melazzi\IEEEauthorrefmark{1},
Felipe Huici\IEEEauthorrefmark{2},
}

\IEEEauthorblockA{
\IEEEauthorrefmark{1}University of Rome Tor Vergata/CNIT,
\IEEEauthorrefmark{2}NEC Laboratories Europe,
\IEEEauthorrefmark{3}Politecnico di Torino,
}

\vspace{1em}

\textit{Extended version of the paper accepted for publication in JSAC special issue on Emerging Technologies in Software-Driven Communication - November 2017}
}

\maketitle

\begin{abstract}
We explore the opportunities and design options enabled by novel SDN and NFV technologies, by re-designing a dynamic Content Delivery Network (CDN) service. Our system, named MOSTO, provides performance levels comparable to that of a regular CDN, but does not require the deployment of a large distributed infrastructure. In the process of designing the system, we identify relevant functions that could be integrated in the future Internet infrastructure. Such functions greatly simplify the design and effectiveness of services such as MOSTO. We demonstrate our system using a mixture of simulation, emulation, testbed experiments and by realizing a proof-of-concept deployment in a planet-wide commercial cloud system.
\end{abstract}


\section{Introduction}
A number of technological trends are transforming the Internet infrastructure.
The fast evolution of \emph{Network Function Virtualization}~\cite{etsi_nfv}, the growing success of initiatives like CORD (Central Office Rearchitected as a Datacenter)~\cite{cord}, the advent of 5G networks~\cite{5g} and Mobile Edge Computing~\cite{hu2015mobile} are instances of this transformation. 
As a result, the envisioned future Internet will provide a big number of locations, distributed in different parts of the network, where new services can be dynamically deployed. We refer to these locations as \emph{Function Execution Points} (FEPs). For instance, 5G networks will provide FEPs in the access networks, Operator's Central Offices will be located in metropolitan aggregation networks, while cloud datacenters and Internet eXchange Points (IXPs) will be close to the network core.

The large availability of FEPs could potentially reshape the way network applications are designed and deployed, as it was the case for cloud datacenters. There, the opportunity to dynamically rent complex computing infrastructures and/or services, from a third-party, has fueled innovation in applications design, and opened the market for new players. 
While we cannot predict the market or future business models, we are interested in shedding some light on the new technical solutions that can be enabled by an infrastructure with a large availability of distributed FEPs. 
To this aim, our approach is to re-design relevant today's network applications from scratch, leveraging such a new infrastructure. 

In this paper, we present the design and implementation of MOSTO, a Content Delivery Network (CDN) for improving the performance of dynamic web content delivery from an origin server. 
The motivation for selecting a CDN as application is twofold. First, CDNs are widely deployed and play a critical role in today's Internet~\cite{mukerjee2016impact}. Second, building a CDN requires a fine engineering work to balance performance, scalability and costs. Complex algorithms for content placement and routing~\cite{Maggs:2015:ANC:2805789.2805800}, fast software implementations~\cite{Kuenzer:2013:TMV:2535828.2535832}, tuning of network protocols~\cite{overclockingYahoo} are just few of the required building blocks, which allow us to touch a large variety of topics. Furthermore, the decision to target dynamic content delivery focuses our discussion on networking-related issues, rather than on cache dynamics as it is the case for static content delivery~\cite{Traverso:2013:TLT:2541468.2541470}.

MOSTO dynamically places virtualized proxies in the network to accelerate TCP flows. In effect, we revisit a well-known technique for TCP acceleration~\cite{appLevelRelays}, i.e., using a chain of TCP proxies to connect the end-points, in the light of recent advances on network function virtualization~\cite{martins2014clickos} and programmable network devices~\cite{bosshart2013forwarding}. 
In particular, we provide two main contributions. First, we describe the way MOSTO addresses the following issues:
\begin{itemize}
    \item Mapping clients to ingress proxies and intercepting client network traffic;
    \item Selecting proxies' locations and proxies chains for a given TCP flow;
    % \item Steering traffic through chains of TCP proxies;
    \item Scaling proxies forwarding throughput to minimize the cost of running the system.
\end{itemize}

\noindent Then, as second contribution, we provide the identification of valuable features future FEPs could provide, and point out how such features could be leveraged in the design of advanced services such as MOSTO. More specifically, we show that: \begin{itemize}
    \item a reactive instantiation strategy of the proxies, together with a very short creation time of the proxy functions, can avoid the need to deploy DNS or anycast traffic redirection strategies;
    \item providing network measurements among FEPs as a service can simplify the design of network applications; 
    \item the ability to program the infrastructure's switches attached to the proxies can reduce the overall system load.
\end{itemize}

To demonstrate and evaluate our system implementation, we rely on a combination of simulation and emulation tests, benchmark tests in a controlled testbed and a proof-of-concept deployment that uses the Amazon EC2 cloud.

% In particular, this paper reports on the way MOSTO addresses the following issues:
% \begin{itemize}
%     \item Mapping clients to ingress proxies and intercepting client network traffic;
%     \item Selecting proxies' locations and proxies chains for a given TCP flow;
%     % \item Steering traffic through chains of TCP proxies;
%     \item Scaling proxies forwarding throughput to minimize the cost of running the system.
% \end{itemize}

% \noindent In the process of describing the implemented solutions and technologies, we identify possible useful future infrastructure features and point out how they could help the rolling out of MOSTO. In particular, we show that a reactive instantiation strategy of the proxies, together with a very short creation time of the proxy function, can avoid the need to deploy DNS or anycast traffic redirection strategies. Furthermore, we show that giving the ability to program the infrastructure's switches attached to the proxies can reduce the overall system load.

% To demonstrate and evaluate our system implementation, we rely on a combination of simulation and emulation tests, on benchmark tests in a controlled testbed and on a proof-of-concept deployment that uses the Amazon EC2 cloud.

\begin{figure*}%
    \vspace{-3ex}

\nocaption

  \centering
  \begin{subfigure}{.47\columnwidth}
    \includegraphics[width=\columnwidth]{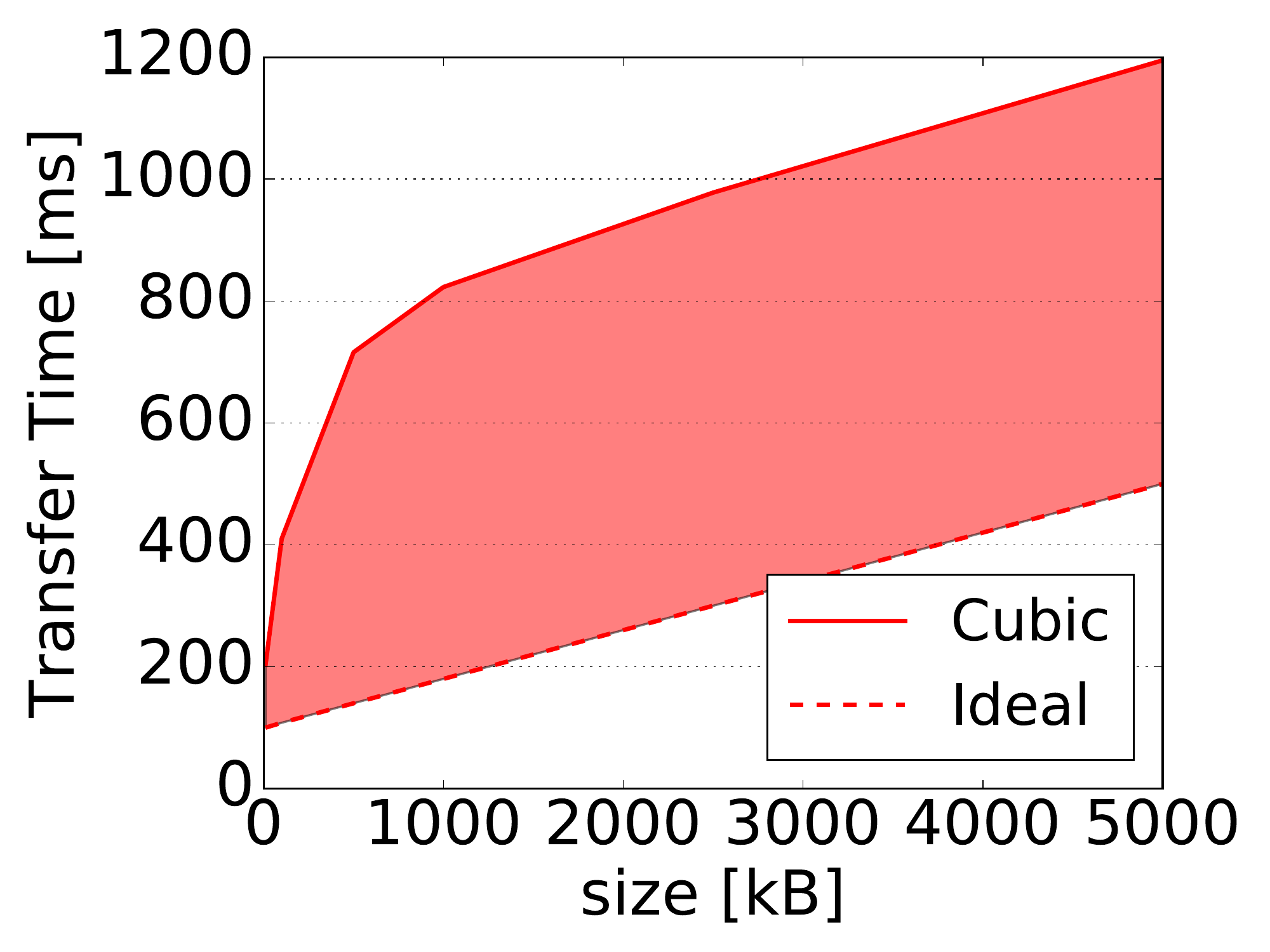}%
    \caption{Ideal and TCP transfer times for different flow sizes. The colored area represent the TCP's overhead.}%
    \label{fig:ideal-gain}%
   \end{subfigure}\hfill%
  \begin{subfigure}{.70\columnwidth}
    \includegraphics[width=\columnwidth]{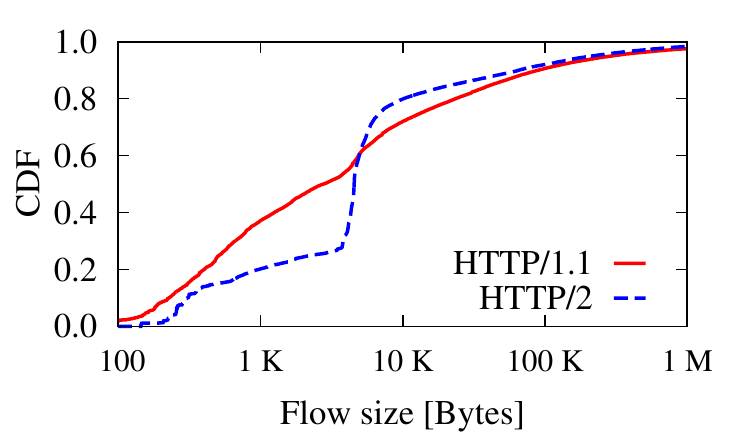}%
    \caption{Variation in flow size caused by HTTP/2 as seen from the vantage point in the Italian operator's PoP.}%
    \label{fig:flowsize2.pdf}%
  \end{subfigure}\hfill%
%   \begin{subfigure}{.65\columnwidth}
%     \includegraphics[width=\columnwidth]{imgs/rttSize}
%     \caption{Distribution of TCP flow sizes and corresponding RTTs. Measurements are taken in a operator's PoP and do not include the last mile delay.}
%     \label{fig:rttSize}
%   \end{subfigure}  % 
%   \caption{Measurement campaign results and impact of TCP mechanics on flow completion time.}
   \begin{subfigure}{.80\columnwidth}
    \includegraphics[width=\columnwidth]{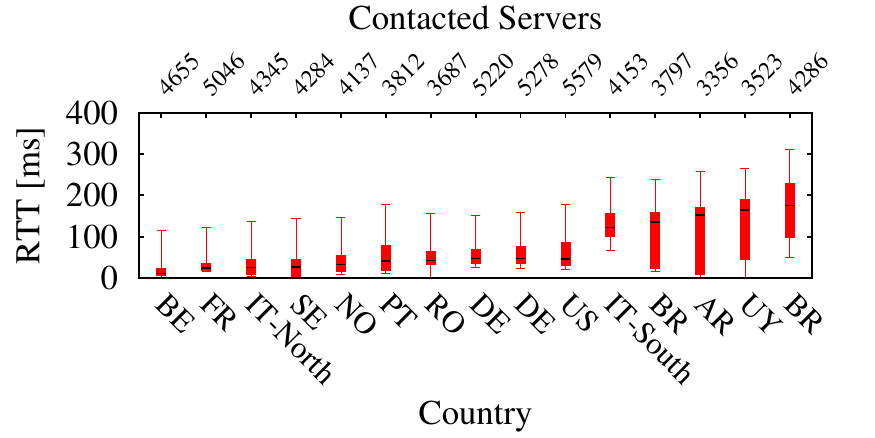}%
    \caption{Distribution of RTTs for the TCP flows used to download the per country Alexa's top 500 websites.}%
    \label{fig:rtt2.pdf}%
   \end{subfigure}
    %\caption{Measurement campaign results and impact of TCP mechanics on flow completion time.}
  \label{fig:measurements}
      \vspace{-2ex}

\end{figure*}
% In fact, while we report on our experience in deploying it in today's Internet using Amazon EC2, we highlight how the inclusion of such features would greatly improve or simplify both MOSTO's design and deployment. 

% The rest of this paper is organized as follows...

The rest of this paper is organized as follows. In Sec.~\ref{sec:backmot} we give a brief description of the problems in dynamic web content delivery, and we present the results of a measurement campaign that motivates our work in optimizing the TCP Slow-Start using MOSTO. Sec.~\ref{sec:mosto} describes MOSTO's architecture and operations, while Sec.~\ref{sec:issues} highlights relevant design issues and related opportunities to leverage the future infrastructure's FEPs. We present an evaluation of our solution in Sec.~\ref{sec:eval}, and discuss In Sec.~\ref{sec:discussion} additional aspects of the MOSTO's design. Sec.~\ref{sec:related} reports on related work and finally in Sec.~\ref{sec:conclusion} we present our conclusions.

\section{Background and motivation}
\label{sec:backmot}

This section presents issues in dynamic web content delivery and briefs on currently employed solutions. Throughout the section we report the results of a measurement campaign
we used to motivate our work and validate our hypothesis.
% therefore, we start by briefly explaining the measurement setup.

\vspace{0.1em}
The performance of dynamic content delivery largely depends on the employed data transfer mechanisms, since strategies such as content caching cannot be employed. For a large portion of Internet's data transfers, the employed mechanisms are those implemented by HTTP and TCP.
In this paper, we focus in particular on the effect of TCP, and specifically on the effect of the \textbf{TCP's Slow-Start algorithm}.
The Slow-Start algorithm forces a connection's end-point to use a slower sending rate during the start-up phase of the connection. 
When a feedback from the other end-point arrives, i.e., after a round-trip time (RTT), the sending rate is increased. When the available bandwidth is finally discovered, the Slow-Start phase ends. The need to wait for a feedback makes the speed of convergence to the available bandwidth related to the RTT between the connection's end-points. 
The effect of such mechanism is presented in Figure \ref{fig:ideal-gain}, which compares the ideal achievable transfer time with the one achieved with a Linux's TCP Cubic implementation, over a 100Mbps link with 100ms RTT. The TCP Cubic Slow-Start limits the sending rate until it converges to the available bandwidth, with a significant increase of the Transfer Time, e.g., 5x for a 500KBs transfer with 100ms RTT.

The performance limitation introduced by Slow-Start has a little impact on today's data transfers, since the majority of TCP flows have a size smaller than 100KBs~\cite{overclockingYahoo}. However, the recent development and deployment of protocols such as SPDY~\cite{spdy} and HTTP/2~\cite{http2} has the potential to change the amount of data transferred by each TCP flow. By multiplexing multiple HTTP requests over a single TCP connection, such protocols effectively enable the use of fewer TCP connections, consequently increasing the amount of data transferred per connection. Such an increase would make TCP transfer times largely dependent on the Slow-Start performance.

To verify this hypothesis, we performed a measurement campaign to understand the distribution of TCP flow sizes and the impact of the deployment of the HTTP/2 protocol. 
In this campaign we collected and analyzed one month of traffic data from a PoP of an Italian ISP in North Italy. The PoP aggregates traffic of more than 8,000 ADSL and FTTH (Fiber-To-The-Home) customers. The data consists in more than 250 GB of (anonymized) log files collected using Tstat~\cite{7876976}, a passive meter that rebuilds each TCP connection and exposes statistics about RTT, packet losses, flow and packet sizes. 
% characterize the correlation between size and RTT of TCP flows and 

Fig.~\ref{fig:flowsize} shows the increase of TCP flow size as measured in the Italian vantage point. Even if the HTTP/2 traffic support is still limited, with only 18\% of the flows being HTTP/2, the average TCP flow size is already clearly larger. Also, we expect the impact of HTTP/2 on the number and size of TCP connections to be significantly more relevant in future. In fact, current web site design best practices are optimized for HTTP/1.1 and usually force the use of multiple TCP connections. On the contrary, best practices for HTTP/2 suggest the use of fewer connections\footnote{For instance, domain sharding, which effectively requires a client to open more TCP connections, was used as a mean to improve web sites performance over HTTP/1 but is a discouraged practice for HTTP/2.}.
% \url{https://blog.cloudflare.com/http-2-for-web-developers/}}. 
As a further demonstration of this proposition, we add that a large content provider's page, e.g., Facebook, which is already optimized for HTTP/2, shows an increase of the average flow size from 45KBs to 210KBs.

\vspace{0.1em}
The TCP Slow-Start issue is particularly important in the context of \textbf{CDNs} and large Content Providers (CPs), where the issue is addressed by deploying huge infrastructures, with many front-end proxies distributed at the network's edge~\cite{googleInfrExp}. A front-end proxy keeps an open TCP connection with the back-end server hosting the content, establishing new connections only with end-users. With this solution, a proxy takes advantage of the proximity to the user to guarantee a very low RTT, reducing the impact of the Slow-Start. However, only such big players can afford the cost of deploying and maintaining large infrastructures. 
Furthermore, while large CPs know in advance the back-end server they need to open TCP connections with, in the case of CDNs the number of different back-end servers could be much larger. In fact, a typical strategy adopted by CDNs is to close a back-end connection after an idle timeout. In effect, such an approach penalizes web sites that are not frequently accessed from a given location. Also, excluding a few CDNs with large footprints, e.g., Akamai, most of the CDNs have a much smaller number of locations. The final outcome is that many users may still experience RTTs of several tens of ms even with CDN front-ends. 

To quantify these effects, we performed a second measurement campaign that analyzes the RTT actually perceived by users of different countries when accessing web content. In particular, we measured the performance of the Alexa per country top 500 web sites from 15 different countries over Europe, North and South America. We deployed our tests on Linux machines at the edge of the network (in ADSL or Fiber installations).
We instrumented Google Chrome to visit the target web pages; the browser downloaded the Alexa top 500 web sites for the target country, 5 runs each.
%Thanks to the developer tools, it was possible to export the HAR (HTTP ARchive \cite{har_spec}) for each visit; it logs all the objects that compose along with metadata (size, timings, headers).
In parallel to Google Chrome, test machines were also running Tstat~\cite{7876976} for collecting detailed flow-level statistics.

Fig. \ref{fig:rtt} shows the distribution of measured RTTs. Boxes embrace quartiles and whiskers the $5^{th}$ and $95^{th}$ percentile; black lines are medians.
The considered websites are likely to use large CDN services to deliver their contents. Nonetheless, a significant fraction of the flows still experiences larger RTTs, of several tens of ms. Also, countries where CDNs have smaller footprints (i.e., South America) show larger mean RTTs~\cite{googleInfrExp}, as expected.
Another observation regards the difference in RTTs experienced by the vantage points in north and south Italy. The south Italian vantage point has large RTTs, likely caused by the peripheral geographic position. In fact, many CDNs deploy caches only in the largely populated area of Milan, which is hundreds of Kms away from southern Italy.

Overall, this measurement campaign provided useful insights about the distribution of RTTs in different locations. Given the impact of additional transfer latencies as low as 100ms~\cite{singla2014internet}, these results confirm that, even in presence of large footprint CDNs, there is still a relevant potential impact of TCP acceleration techniques, such as those that we are proposing with MOSTO.

\section{MOSTO}
\label{sec:mosto}

% In this paper, we try to move a step forward in the understanding of the potential impact such a future Internet infrastructure could have on network applications.

% we explore the effect of this new infrastructure on one popular network service: content distribution.

% Considering the potentially big number of newly available FEPs, their decentralized nature, and the generally smaller scale of available resources compared to large cloud datacenters, we also expect these FEPs to provide a set of helper services to orchestrate and manage the deployed third-party services. For instance, we expect something similar to the Amazon AWS's \emph{lambda} services, where the instantiation and execution of a function is triggered by an event, instead of being explicitly triggered by a centralized management system.

% Achieving efficient data transfers is an overarching goal of any communication system. Network delay is a critical performance metric for many applications~\cite{speedOfLight}.

In this section, we present the architecture and operations of MOSTO (Managed Overlay for Slow-starT  Optimization). 
MOSTO accelerates the delivery of dynamic (i.e. non-cachable) web content, providing a performance level similar to the one achieved with a widely spread, regular CDN infrastructure that pre-establishes back-end connections. However, contrary to regular CDNs, MOSTO does not require a widely distributed infrastructure to be deployed in advance.

% , a system to improve the Slow-Start performance for selected TCP flows.

In a nutshell, MOSTO uses a managed peer-to-peer overlay to create a chain of TCP proxies between two end-points, accelerating the convergence to the available bandwidth during the TCP Slow-Start phase. While the  approach is not new and has been already studied in the past~\cite{appLevelRelays}, our solution introduces novel aspects tackling relevant deployment and scalability issues. MOSTO has two main advantages when compared with state of the art CDNs. First, it is does not require the proactive deployment of a large infrastructure. Second, it can be faster than a regular CDN, since MOSTO can place proxies in locations where a deployment of CDN nodes is difficult. 
% Third, MOSTO can be completely transparent to users (and to servers) since it only operates at the transport layer: a feature particularly useful when dealing with encrypted content.

\subsection{Architecture}

\begin{figure}
    \includegraphics[width=\columnwidth]{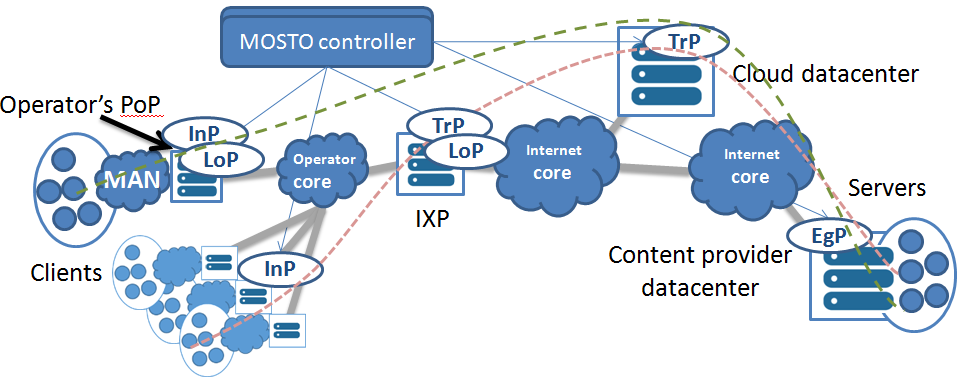}%
    \caption{MOSTO architecture}%
    \label{fig:mosto}%
    \vspace{-2ex}

\end{figure}

MOSTO (Fig. \ref{fig:mosto}) is composed of a network of TCP proxies and of a centralized controller that manages them.  
For the deployment of the proxies, we assume that a number of FEPs are distributed at different network locations and can run third party services, e.g., in the form of Virtual Machines (VMs) or containers. We also assume these locations to be typically placed close to the transit points of many network flows.

In the most general case, there are three different types of proxies in the overlay. The  overlay entry and exit points are named ingress proxy (InP) and egress proxy (EgP), respectively. The proxies used to connect the InP to the EgP are named transport proxies (TrPs). Furthermore, either the InP or one of the TrPs can play the additional role of a lookup proxy (LoP). The LoP is in charge of selecting the right chain of proxies that should be used for a given TCP flow. 
% A TCP proxy in MOSTO can play one or more of the above roles. For instance, very often an InP is also a LoP.

%  The  decision  on  where  perform the lookup operation depends on the deploy location used by the  chain.  If  the  ingress  node  is  located  in  a  data  center  that can  potentially  serve  a  large  number  of  users,  this  node  will perform the look up. If the ingress node is placed in a Flexible Location that serve a limited number of users, or even a singl euser  (e.g.  home  gateway  or  DSLAM)  the  lookup  operation will be delegated to an aggregation node. The node that will perform the lookup receives from the controller node constant updates:  about  the  network  state  and  about  which  chain  use depending on which content is requested. For these reasons is convinient to limit the number of node that perform the lookup and  place  them  in  location  where  they  can  serve  the  highest number of users.

The MOSTO controller is in charge of selecting proxies' locations and of performing orchestration and management of the infrastructure. It runs an optimization algorithm designed to maximize TCP performance. The algorithm selects the number of proxies to be used per connection and their locations in order to minimize the overall RTT, while evenly splitting the delays between proxies. This has been proven optimal for the TCP connection performance~\cite{appLevelRelays}.

The MOSTO control plane has to perform three operations whenever a new TCP flow starts.
First, it has to select the InP for the new TCP flow. For instance, state-of-the art strategies like DNS-redirection can be applied here. 
% However, MOSTO implements a different approach, which is discussed in Sec.~\ref{sec:endUsrMapping}. 
Second, the MOSTO control plane needs to select an EgP. While different approaches can be used, throughout the paper we assume that the closest location to the TCP flow destination is selected to host the proxy. This assignment can be performed both statically, e.g., using the geographical position to compute the distance, or dynamically, e.g., using periodic active measurements from different possible proxy locations.
Finally, MOSTO computes an optimal proxy-to-proxy chain between InP and EgP. We describe in details this operation in the next paragraph.

% For the deployment of the proxies, we assume that a number of network's locations offer the possibility to run third party services, e.g., in the form of Virtual Machines (VMs). More specifically, in addition to today's cloud datacenters, we envision also locations such as virtualized carrier's \textit{Point-of-Presence}s (PoPs)~\cite{cord}, Internet Exchange Points (IXPs)~\cite{IXP}, Mobile Edge-Computing infrastructures~\cite{?} to name a few. 

\subsection{Computing proxy-to-proxy chains}
The MOSTO controller is in charge of selecting the proxy chain that maximizes performance for a given TCP flow. This is a challenge, considering the need of taking a decision on-the-fly (i.e., in few ms) when a new flow starts.

\vspace{0.2em}
\noindent\textbf{Problem definition} 
We model the chain selection as an optimization problem. In particular, we focus only on the TCP Slow-Start performance, assuming that bandwidth is reasonably large between proxies. Here, notice that the last mile latency is the usual bottleneck in terms of throughput~\cite{Sundaresan:2011:BIP:2018436.2018452}. Furthermore, MOSTO optimizes short transfers, for which the RTT, and not the bandwidth, is the dominant performance factor~\cite{singla2014internet}.

For a given proxy path, the TCP transfer time during Slow-Start is modeled as a linear combination of the path length (i.e., sum of RTTs) and the maximum RTT within the path, i.e., between any pair of proxies~\cite{siracusano2016fly}. The coefficients of this linear combination depend on the size of the data transfer, hence, different transfer sizes can have different optimal paths. 
We denote by the \emph{distance} $d(p,q)$ the packet RTT between proxies~$p$ and~$q$. The time required for a data transfer consisting of $r$ rounds across a given proxy path $p_1,\ldots,p_k$ calculates as
$$
	\frac{1}{2} \cdot \sum_{i=1}^{k-1} d(p_i,p_{i+1}) + r-1 \cdot \max_{i=1}^{k-1} d(p_i,p_{i+1}) \ ,
$$
where the first summand represents the time from initiation of the data transfer at $p_1$ until the packets of the first round arrive at $p_k$.  The second summand is, $r-1$ times, the time passing between the packets of two successive rounds to arrive at $p_k$. This inter-arrival time at the last proxy is determined by the maximum RTT between any proxy pair on the path. In fact, the maximum inter-proxies RTT will limit the end-to-end throughput given the back-pressure behavior of a TCP connection.

\vspace{0.2em}
\noindent\textbf{Algorithm}
The MOSTO controller splits the path computation operation in two: pre-computation and lookup. 
% This split allows MOSTO to perform a very fast online selection of the path for an incoming TCP flow.

\vspace{0.1em}
\noindent\textit{Pre-computation:}
taking as input all the RTTs between any pair of proxy locations, this step provides the complete set of possible optimal paths between any pair of proxies, for several transfer size\footnote{The actual transfer sizes can be grouped in discrete intervals given by the successive increments of the TCP congestion window.}. 
We implemented the pre-computation using an extension of the well-known Floyd–Warshall dynamic programming algorithm for the all-pair-shortest-path problem.
% , using an additional technique to avoid unnecessary computations. 
The complexity for this computation is $\Theta(n^2 m)$, where $n$ is the number of possible proxy locations and $m$ is the number of direct proxy-to-proxy links in the proxy network. Since we consider TCP proxies, they are connected in a full mesh overlay network, i.e., each proxy can directly send packets to any other proxy. Therefore, the complexity can be also expressed as $\Theta(n^2 m) = \Theta(n^4)$.
% , because for each such combination we need to apply the recursive formula or a base case once. 
To speed-up the process and increase controller scalability, our implementation realizes an optimization technique that avoids unnecessary computations. Both the dynamic programming algorithm and the mathematical observations that enable the mentioned optimization are described and demonstrated in appendix to this paper. Here, we point to the fact that the final controller algorithm is still providing a complexity $O(n^4)$ in the worst case, but regular case and best case complexities are improved. For instance, the best case is improved to $\Omega(n^2)$. The actual runtime in practice depends on the graph structure and it is experimentally evaluated in Sec.~\ref{sec:pathselection}.

\vspace{0.1em}
\noindent\textit{Lookup:}  
The pre-computation algorithm provides, for each proxy pair, the set of all Pareto-optimal paths in terms of the two metrics (path length and maximum distance between the proxies). Once the pre-computation is done, determining the best path for a data transfer is a matter of performing a lookup in a table.
The lookup operation obtains the optimal path using a InP-EgP couple and the transfer size as lookup key. 

\vspace{0.2em}
\noindent\textbf{Realization}
The pre-computation, which is computationally expensive and takes a longer time, can be performed offline by the controller. In fact, the possible proxy locations are known in advance to the controller, since they correspond to FEPs. This pre-computation has to be performed periodically, since it takes as input a distance matrix built by measuring the RTTs between all proxy pairs. The periodicity depends on the speed at which the RTTs change. Related work suggests that performing the pre-computation every few minutes is sufficient to capture significant RTT changes~\cite{bifulco2015fingerprinting}. We empirically fixed such period to 5min. Such periodicity also requires that the pre-computation should complete in few seconds in order to provide timely updated solutions. The pre-computed paths are then distributed to the proxies (LoPs), which perform the path lookup.

% In more general terms, for a given proxy path the transfer time is a linear combination of the path length (i.e. sum of distances) and the maximum distance among the path. As a consequence, there is no unique best path between a given pair of proxies, because the coefficients of the linear combination depend on the size of the data transfer in the way described in the above formula, and thus different transfer sizes can have different optimal paths. To deal with this situation in our pre-computation algorithm, we compute for each proxy pair the set of all Pareto-optimal paths in terms of the two metrics (path length and maximum distance). This Pareto-front contains for any transfer size the optimal path.

\subsection{Operations overview}
% As it happens with CDNs, 
% Having described the system components in the previous paragraphs, we now provide an overview of its operations. 

When a user starts a new connection with a web site delivered through MOSTO, the user's network flow is intercepted by a MOSTO InP.
We call \textit{end-user mapping and redirection} the problem of assigning and steering a user's network flow to its ingress proxy.

If the InP is also a LoP, then it performs the lookup function that associates a given destination with a chain of proxies. Otherwise, the InP just forwards the flow to the (statically selected) closest LoP, which finally looks up the chain to be applied. The chains of proxies among which a LoP performs the selection are periodically provided by the MOSTO controller. As discussed in the previous section, the controller needs to collect network measurements between any pairs of locations to perform the pre-computation of the chains. We call \textit{network performance discovery} the problem of collecting network measurements between location pairs. 

% \todo{why we measure only RRT? Congestion does not count because we target short flows (not true since all flows are redirected), RTT gives also a good indication of congestion??. Probably the most safe assumption is that we don't care for bandwidth a loss since we assume connectivity is good between any location pairs, but we can take it into account in the controller if we want to enforce load balancing}

Finally, if a network flow does not end during its Slow-Start phase, then MOSTO gradually removes the proxies from the end-to-end connection. Once a proxy is removed, the connection is again directly operated by the two end-points, reducing the overall load on the MOSTO proxies.
We call this operation \textit{proxy offloading}.

\section{Issues and opportunities}
\label{sec:issues}
In the previous sub-section, we explicitly named three issues, out of several others, which need to be addressed for the design of MOSTO: \textit{end-user mapping and redirection; network performance  discovery; proxy offloading}. We now analyze in more detail such issues and discuss how it is possible to improve the related performance by adding few functions in the future Internet virtualized infrastructure, i.e., in the FEPs.

% \vspace{0.2em}
% \noindent\textbf{
\subsection{End-user mapping and redirection}

\vspace{0.1em}
\noindent\textbf{State-of-the-art}
A network flow can be redirected to an InP using state-of-the-art techniques such as DNS redirection or IP anycast. 
% \vspace{0.1in}
% \noindent\textbf{Issues}
However, building an effective DNS redirection system is challenging and expensive~\cite{chen2015end}. IP anycast is a simpler alternative, but it offers sub-optimal performance~\cite{calder2015analyzing}.

\vspace{0.1em}
\noindent\textbf{Opportunity}
Future networks are expected to offer a large number of FEPs at the very edge of the network~\cite{cord, swBRAS}, pretty close to users. These FEPs are actually on the network traffic's path, e.g., they are co-located with access networks' gateways~\cite{ruckert2014flexible}. A proxy placed in one of this FEPs would not require the enforcement of any redirection or traffic steering on a global scale, because the proxy would be already on path. 

\vspace{0.1em}
\noindent\textbf{Envisioned solution}
Deploying proxies in advance to all the possible edge FEPs is not a scalable solution. First, edge locations, e.g., carrier's central offices, are typically numbered in hundreds per country, which would require the proactive deployment of several thousand proxies to have a large coverage. Furthermore, it would require managing and running proxies even in locations that do not see the traffic of any user. Second, from an FEP provider perspective, it would require hosting a large number of proxies (and services, in general), from many different tenants, in locations that have limited hosting capabilities (e.g., few tens of servers).  

Instead, a viable solution would be to run a proxy only when there are network flows that have to be accelerated using MOSTO. In particular, we envision a FEP to provide a set of helper services to orchestrate the deployed third-party services. This could be similar to the \emph{lambda} services in Amazon AWS, where the instantiation and execution of a function is triggered by an event, instead of being explicitly controlled by a centralized management system. Realizing such a system requires the definition of a \textit{triggering mechanism} and the realization of \textit{fast booting} service instances.

\vspace{0.1em}
\noindent\textit{Triggering:} we trigger the creation of a proxy in a FEP whenever the FEP is on the path of a network flow that has to be handled by MOSTO.
Actually, the deployment process requires two steps. First, the registration of a \emph{forwarding entry} in the FEP, to match the traffic directed to MOSTO and forward it towards the proxy. Second, the provisioning of a proxy \emph{image} to be executed whenever there is no other MOSTO proxy already running, and a new flow is matched by the forwarding entry.
Of course, we also expect mechanisms to automatically terminate proxies when the network flows become inactive, and to auto-scale the number of proxies with the network load. These mechanisms are however already envisioned in architectures for NFV~\cite{etsi_nfv}.

\vspace{0.1em}
\noindent\textit{Fast booting:} for the solution to be effective, a MOSTO proxy has to boot in few ms. In fact, using mechanisms such as the one envisioned in \cite{jitsu}, one can boot a TCP proxy during the time required to perform the TCP's three-way handshake. More specifically, once a TCP SYN is received, the triggering mechanism generates in response a TCP SYN|ACK, while at the same time it starts the proxy. Then, once ready, the TCP connection state is handed to the proxy\footnote{The TCP state at this point includes minimal information, such as the selected TCP sequence numbers.}. We implemented this mechanism using Miniproxy~\cite{siracusano2016fly}, a unikernel~\cite{unikernels} implementation of a TCP proxy, which can boot in about 10ms.

% \vspace{0.2em}
% \noindent\textbf{
\subsection{Network performance discovery}

\vspace{0.1em}
\noindent\textbf{State-of-the-art}
The MOSTO controller takes a distance matrix as input for the computation of the proxy chains that should be used for accelerating the connection between any pair of end-points. Such distance matrix contains the distance, in terms of RTT, between any pair of possible locations where proxies can be run. That is, the RTTs between any pair of FEPs.
Current CDN deployments also require such information and hence rely on a continuous network monitoring infrastructure in which each CDN proxy periodically measures the network performance on the path towards the other proxies~\cite{chandrasekaran2015server}.

\vspace{0.1em}
\noindent\textbf{Opportunity}
Such a measurement infrastructure is not an exclusive need of CDN services. In fact, several services have a similar requirement~\cite{via,Su2006SIGCOMM,microsoftInfr,Gummadi2004OSDI}. Therefore, FEPs may provide such information as a service, simplifying the design of applications that require such information, while optimizing the FEP and network resources usage. In fact, the FEP can perform the measurements once and for all the interested parties.

\vspace{0.1em}
\noindent\textbf{Envisioned solution}
We envision that FEPs offer a service to register \textit{measurement requests} for the performance of paths to other FEPs. Likewise, a FEP  offers a measurement end-point that other FEPs can, e.g., \textit{ping} to measure network performance. 

In realizing the MOSTO controller, we assume that such a service is available. Furthermore, we assume that a FEP provides updates to the MOSTO controller only when a given measured value, e.g., the RTT, changes of a value bigger than a given threshold. Thus, the MOSTO controller registers a number of measurement requests to the interested FEPs. The selection of which requests to register, at which FEPs, will largely depend on the network structure and on the service cost. For example, locations which are far away from each other may not need to measure the network performance among them, saving on the number of registered measurements requests.

\begin{figure*} [!ht]%
    \vspace{-3ex}

  \centering
  \begin{subfigure}{.55\columnwidth}
    \includegraphics[width=\columnwidth]{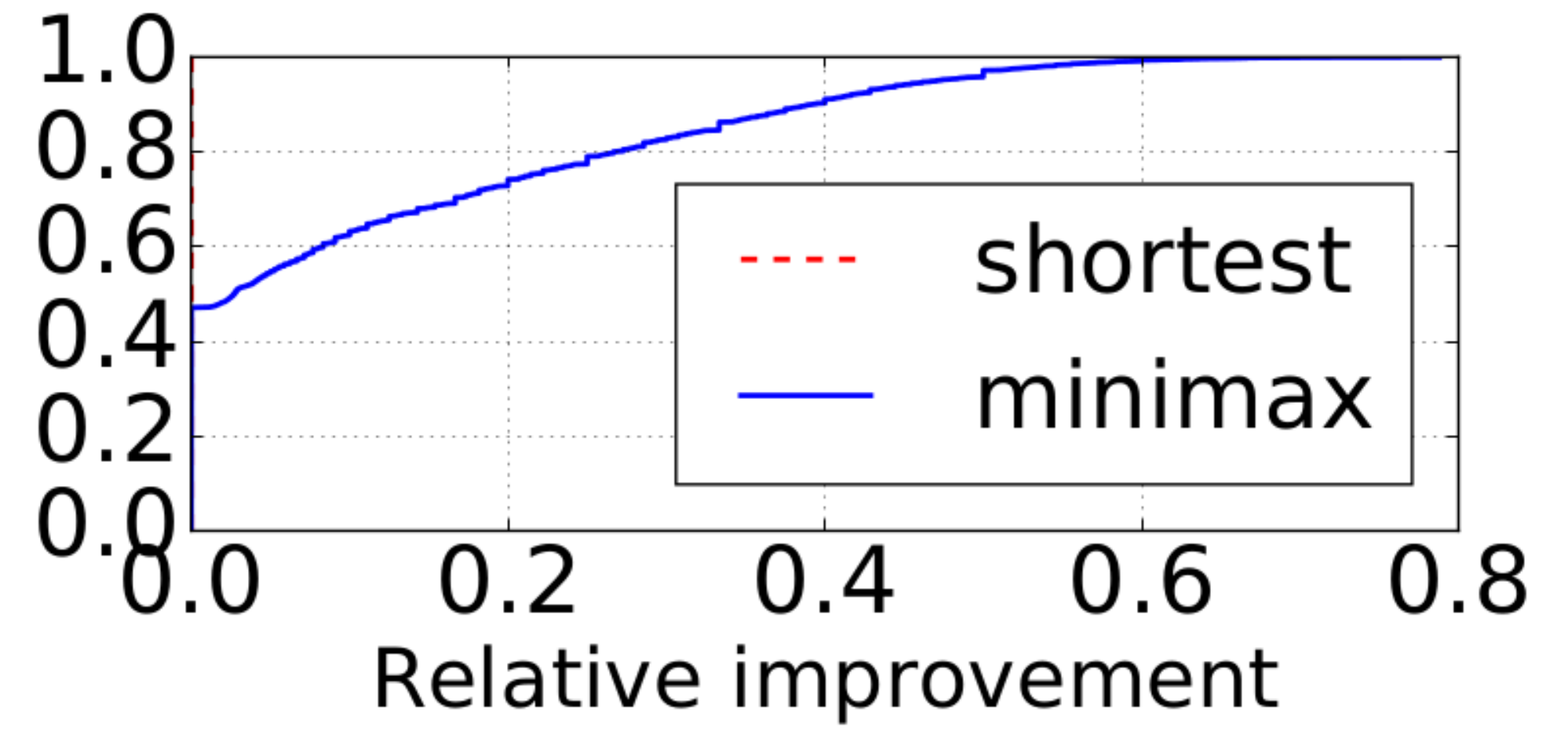}%
    \caption{$<$15KBs (1 round)}%
    \label{fig:k1cdf}%
  \end{subfigure}\hfill%
  \begin{subfigure}{.55\columnwidth}
    \includegraphics[width=\columnwidth]{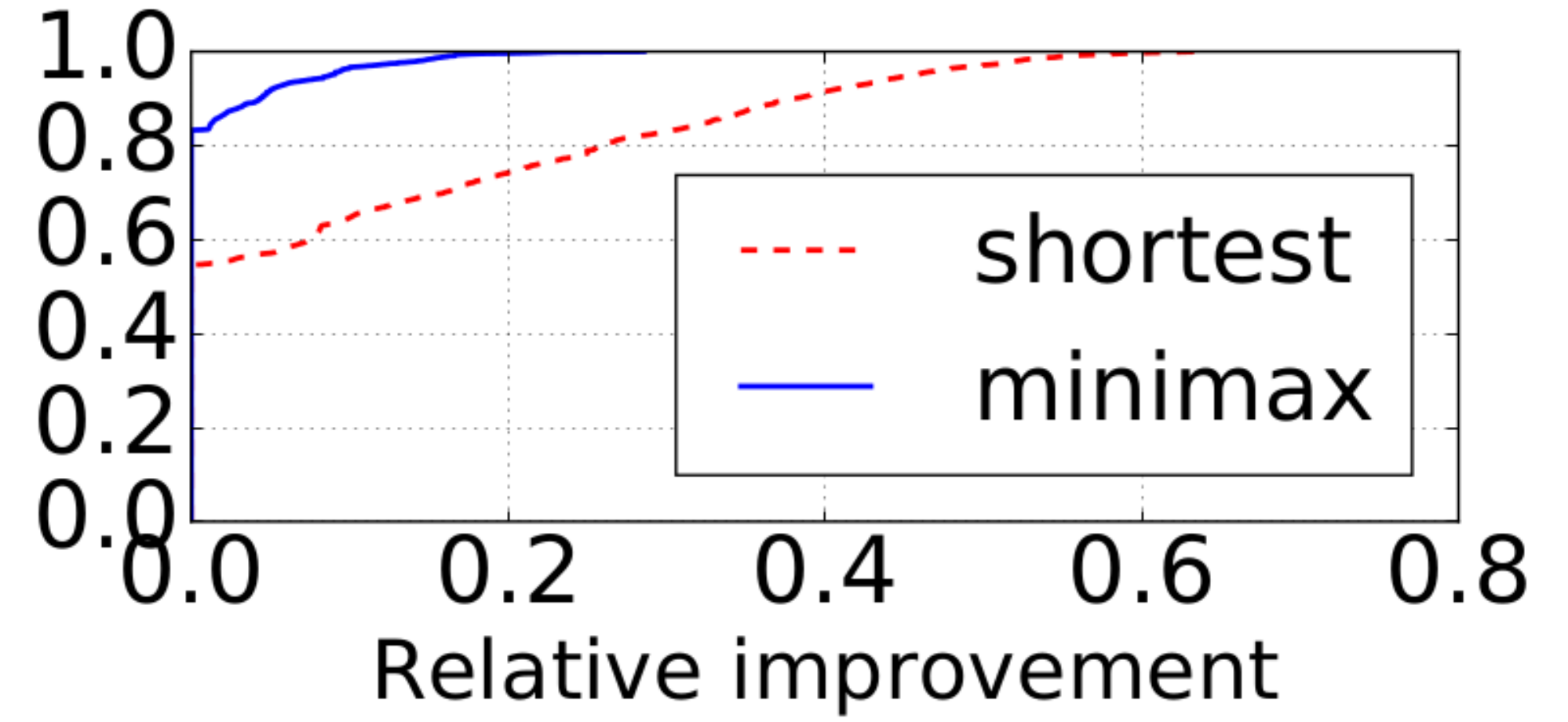}%
    \caption{450KBs (5 rounds)}%
    \label{fig:k5cdf}%
   \end{subfigure}\hfill%
  \begin{subfigure}{.55\columnwidth}
    \includegraphics[width=\columnwidth]{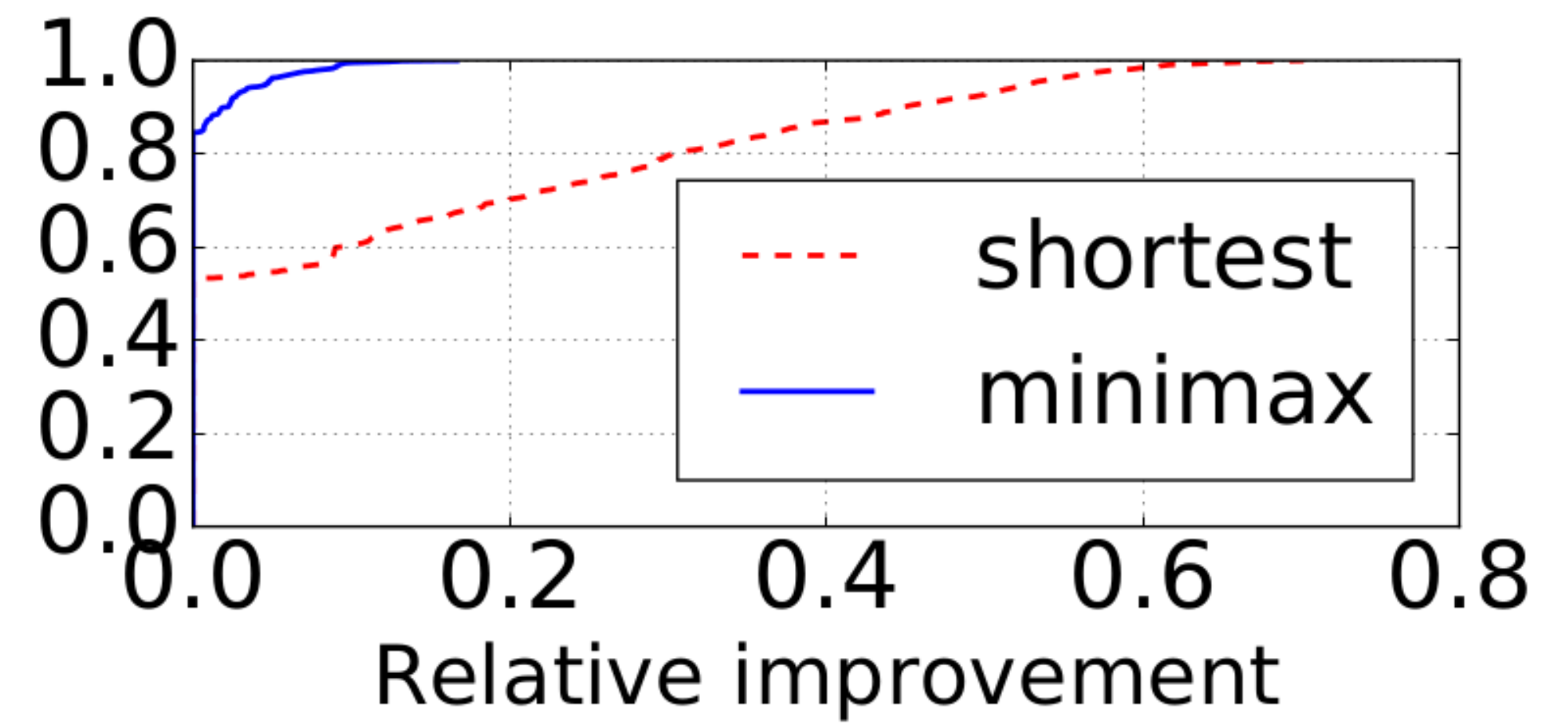}%
    \caption{14935KBs (10 rounds)}%
    \label{fig:k10cdf}%
   \end{subfigure}  % 
  \caption{Improvement over shortest or minimax path selection for different packet sizes}
  \label{fig:pathselperf}
      \vspace{-2ex}

\end{figure*}

% \vspace{0.2em}
% \noindent\textbf{

% \vspace{0.2em}
% \noindent\textbf{
\subsection{Proxy offloading}
\vspace{0.1em}
\noindent\textbf{State-of-the-art}
Once the TCP Slow-Start is over, the benefit of proxies on the TCP flow performance is limited\footnote{Actually, in presence of significant packet loss, a chain of proxies would still provide benefits for many TCP implementations.}. To minimize the cost of processing network flows, it is therefore convenient to remove a proxy from the end-to-end path. Unfortunately it is not possible to completely remove a proxy from an active chain, since two different TCP connections would need to be joined into one. Most notably, while the congestion control loop can be performed by the joined connections' end-points, such connections will have different sequence numbers to identify the TCP segments, therefore a translation function for the sequence number is still required within the network. For this reason, in our design a MOSTO proxy only performs sequence number translation once the Slow-Start ends, hence reducing the processing load.

\vspace{0.1em}
\noindent\textbf{Opportunity}
While this approach already increases the efficiency of a proxy, the sequence number translation could be performed in the context of the packet switching operations performed by the node running the MOSTO proxy. This will completely remove the need to hand over the packet to the MOSTO TCP proxy once the Slow-Start ends. In particular, the sequence number translation can be performed by programmable SDN switches, such as OpenFlow ones~\cite{mckeown2008openflow}, or fully programmable switches, such as RMT~\cite{bosshart2013forwarding}. 

\vspace{0.1em}
\noindent\textbf{Envisioned solution}
We observe that such a programmable switch could be already deployed in a FEP, and that most likely it is already handling traffic to/from the proxy. Therefore, if the FEP would offer the possibility to program the forwarding action for the entries that handle the proxy's traffic, the MOSTO proxy would be completely offloaded of a particular network flow. At the same time, the switch would still handle the same amount of traffic, with little overhead, if at all, to perform the programmed forwarding action.

Our approach is to program a switch co-located with the proxy to perform the required TCP sequence number translation. We use the P4 language~\cite{bosshart2014p4} to define the sequence number translation action, and PISCES for the switch implementation~\cite{shahbaz2016pisces}. PISCES compiles a P4 switch data plane description in an OpenVSwitch-like software switch implementation.

The MOSTO proxy triggers the insertion of a forwarding entry in the switch, to perform the sequence number translation, when the Slow-Start of a connection ends. More specifically, the proxy waits for the data buffers of both TCP connections to be empty, before actually installing the entry. In fact, empty buffers correspond to no in-flight data. Thus, no retransmission will be required at any later point in time, which makes the switch able to fully take over the connections' packet forwarding.

\section{Implementation and evaluation}
\label{sec:eval}
%\begin{figure*} [!ht]%
%  \centering
%  \begin{subfigure}{.55\columnwidth}
%    \includegraphics[width=\columnwidth]{imgs/K1_cdf.pdf}%
%    \caption{$<$15KBs (1 round)}%
%    \label{fig:k1cdf}%
%  \end{subfigure}\hfill%
%  \begin{subfigure}{.55\columnwidth}
%    \includegraphics[width=\columnwidth]{imgs/K5_cdf.pdf}%
%    \caption{450KBs (5 rounds)}%
%    \label{fig:k5cdf}%
%   \end{subfigure}\hfill%
%  \begin{subfigure}{.55\columnwidth}
%    \includegraphics[width=\columnwidth]{imgs/K10_cdf.pdf}%
%    \caption{14935KBs (10 rounds)}%
%    \label{fig:k10cdf}%
%   \end{subfigure}  % 
%  \caption{Improvement over shortest or minimax path selection for different packet sizes}
%  \label{fig:pathselperf}
%\end{figure*}

%This section presents MOSTO's implementation details, an evaluation of system components, such as the controller path computation algorithm, the proxy offloading mechanism, and the performance achieved by a proof-of-concept deployment that uses the Amazon cloud. 

This section reports the details of the MOSTO implementation and discusses its evaluation. In particular, we present an evaluation of key systems components, such as the controller path computation algorithm and the proxy offloading mechanism. Then, we present a proof-of-concept deployment that uses the Amazon cloud and its performance evaluation.

% In particular, we demonstrate the applicability and effectiveness of the system presenting the results of both testbed measurements and test deployments that use the Amazon cloud to run the proxies.

\subsection{Path selection}
\label{sec:pathselection}
The MOSTO controller is implemented in python, relying on optimized C libraries to perform the path computation.

We evaluated the path computation runtime, the quality of the produced solutions and the number of proxies on the selected paths. In all cases, we run the path computation algorithm on graphs generated using Rocketfuel~\cite{mahajan2002inferring} topologies, which also provide per-link delays. We assume each node in the graph to be a potential location for running a proxy. To build a full mesh network, we introduce links between non-adjacent locations, computing the delay as the sum of link's delays on the shortest path between such locations.  In total, we generate 54 graphs with a number of nodes varying between 7 and 115.

\vspace{0.1em}
\noindent\textbf{Runtime}
For the largest topology of 115 nodes (and about 13k links) our algorithm could compute all the paths, between any pair of proxy locations, in less than one second. A runtime that easily meets our need to execute the algorithm every 5 minutes. Here, it is important to notice that the MOSTO's architectural split between InP and LoP helps in keeping the problem complexity low. In fact, for many InPs located at the very edge of the network, there is only one possible link that can be used by the network flows. For such InPs, MOSTO statically configures the first part of the proxy chain, steering the network flows to the closest LoP, and therefore reducing the number of nodes and links that have to be considered during the path computation.

\vspace{0.1em}
\noindent\textbf{Solution quality}
We compared the computed solutions, for each pair of locations, to the solutions that would have been provided by a shortest path algorithm and by a minimax path algorithm. Here, notice that a shortest path algorithm is preferred when the transfer size is very small, since it optimizes the end-to-end delay. A minimax path algorithm, instead, minimizes the maximum delay on the single links, without caring about the total end-to-end delay. Therefore, a minimax path algorithm generally provides better solutions for larger transfer sizes, since it makes the Slow-Start quicker. 

In Fig.~\ref{fig:pathselperf} we show the results of this test for different transfer sizes. The figure shows Cumulative Distribution Functions (CDFs) of the relative improvement our algorithm's solution provides when compared to the other algorithms' solutions, for each pair of nodes and for each of the considered graphs. 
Fig.\ref{fig:k1cdf} shows the result for a transfer size smaller than 15KBs, i.e., flows that complete the transfer within the initial TCP congestion window\footnote{We assume TCP's ICW=10~\cite{chu2013increasing}.}. In this case, the shortest path algorithm provides the best possible solution, which is the same selected by our algorithm. A minimax path algorithm, instead, is providing worse solutions, with 10\% of the solutions experiencing more than 40\% increase in the transfer time, when compared to the optimal solution.

Fig.\ref{fig:k5cdf} and Fig.\ref{fig:k10cdf} show the result for larger transfer sizes, which require 5 and 10 rounds to complete, respectively. Here, we can see that a shortest path algorithm is not effective, while the minimax path is in many cases provides a good solution. However, in almost 20\% of the cases, our algorithm's solutions still provide an improvement up to the 15\% and 10\% for transfers that take 5 and 10 rounds, respectively.  

\vspace{0.1em}
\noindent\textbf{Number of proxies}
In the process of choosing a path, our algorithm selects also the minimal number of required proxies to provide the optimal solution. As a reference, our algorithm builds solution which on average use 15\% and 14\% less proxies, when compared to a minimax path solution for the cases of 5 and 10 rounds, respectively.

% \vspace{-2ex}

\subsection{TCP Proxy and offloading}

We build MOSTO proxies using Miniproxy~\cite{siracusano2016fly}, a Xen~\cite{barham2003xen} unikernel that implements a high-performance, lightweight TCP proxy. The details of Miniproxy's implementation are presented in \cite{siracusano2016fly}, here we just report that it boots in 12ms and requires as little as 6MBs of RAM to run. For the implementation of MOSTO, we added support for TCP segmentation offloading to the Miniproxy's network stack, achieving line rate 10Gbps throughput on a single core when proxying TCP connections\footnote{If not differently stated, all the tests were performed on an Intel Xeon Ivy Bridge CPU@3.4GHz, 16GB RAM, with a dual port Intel x450 10Gb NIC, running Xen 4.4. A similar server works as traffic generator and receiver. The two servers are connected back-to-back, using the two NIC's port, i.e., creating a bi-directional 10Gbps pipe.}.

To support the offloading to a programmable switch, we further extended Miniproxy to send forwarding entries configuration commands to a software switch.
Our modified proxy monitors the TCP connections' sending rate to detect the end of the Slow-Start. Then, it waits for the receive and sending buffers of the proxied connection to be empty. Once all the outstanding ACKs have been generated or received, the proxy triggers the installation of a forwarding entry, which modifies TCP sequence numbers, in the software switch.

We used a slightly modified version of PISCES~\cite{shahbaz2016pisces} to implement a programmable TCP sequence number \textit{translation} action\footnote{In effect, the sequence number translation corresponds to add/subtract a value to a packet's TCP sequence number and recompute relevant checksums.}. We described such action, and the forwarding tables that make use of it, using the P4~\cite{bosshart2014p4} language.

\vspace{0.1em}
\noindent
\textbf{Throughput}
We performed a throughput test establishing a connection through a MOSTO proxy running on a single CPU's core. Once we reach 100\% utilization of the proxy's CPU core, we measure the achieved throughput and then start the offloading of the TCP connection to the switch. Once the connection is offloaded, we measure again the maximum achieved throughput.
Our results show that in both cases we achieve line rate 10Gbps. 

\vspace{0.1em}
\noindent
\textbf{System load}
When the forwarding is offloaded to the switch, the proxy's CPU core stays idle, while the switch resource consumption is not increased. In fact, the switch is actually experiencing lower load. More specifically, recall that the switch has to forward a received packet twice: a first time to move the packet from the Network Interface Card (NIC) to the proxy; and a second time to move the packet from the proxy to the NIC. When the proxy offloads a connection, the corresponding packets is instead forwarded only once, from NIC to NIC. Also, the cost of performing the TCP sequence number translation action on a packet is negligible, when compared to the cost of performing packet parsing and forwarding table's lookup operations~\cite{softflow}. Thus, in presence of offloading, the system saves both proxy-related processing, i.e., one CPU's core, and one packet forwarding operation, per each received packet, at the switch, while still delivering 10Gbps.

%\begin{figure}
%    \includegraphics[width=\columnwidth]{imgs/offloaded_vs_proxy.pdf}%
%    \caption{Offloading performance}%
%    \label{fig:throughput}%
%\end{figure}

\begin{figure}
    %\vspace{-3ex}

    \includegraphics[width=\columnwidth]{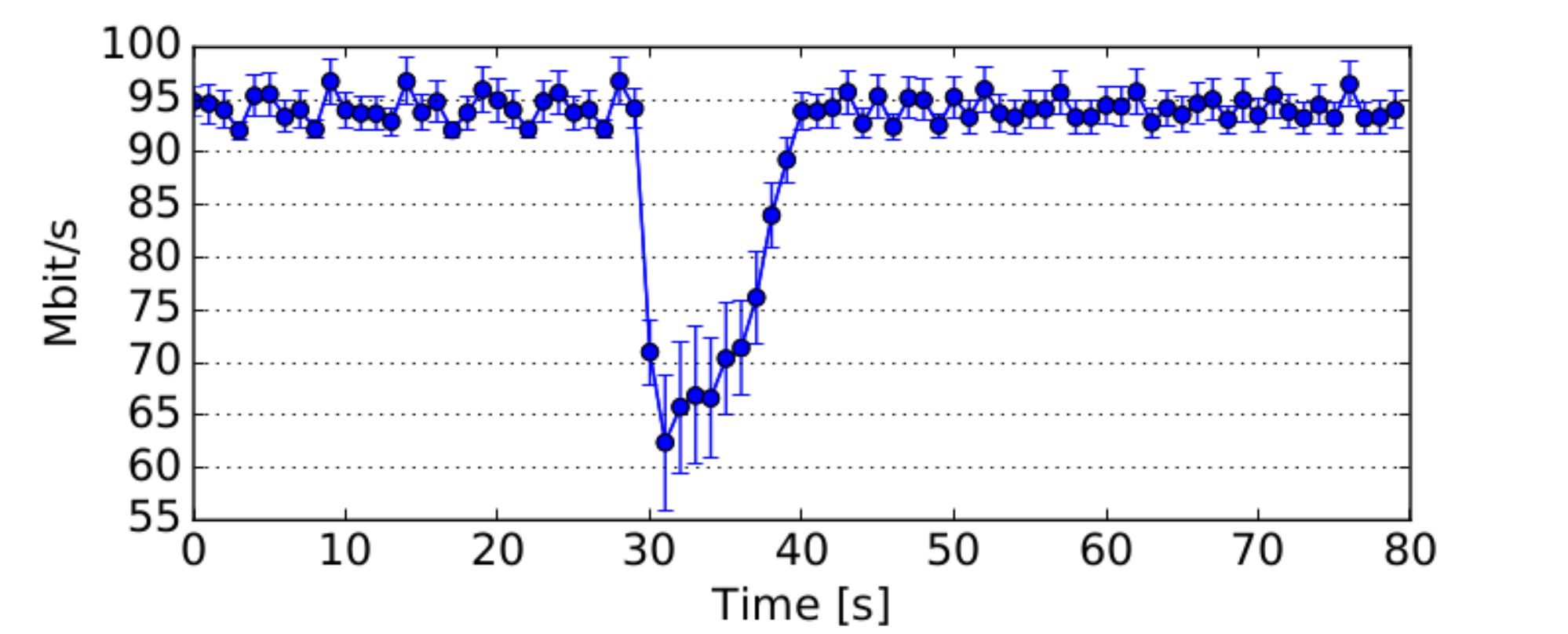}%
    \caption{Offloading effect, varing RTT from 50 to 100 ms}%
    \label{fig:offloadingEffect}%
    \vspace{-2ex}

\end{figure}

\vspace{0.1em}
\noindent
\textbf{Impact on congestion control}
Offloading a connection to the switch actually removes the proxy from the connection's congestion control loop, which is then completely handled by the joined connection's end-points. At the end-points, the removal of the proxy corresponds to a sudden variation in the measured end-to-end delay. Such variation may harm the connection performance, since it may trigger TCP's retransmission timeouts. Therefore, the congestion control loop may wrongly think that a packet not yet received is lost. Such an effect is presented in Fig.\ref{fig:offloadingEffect}. The throughput of an established connection suddenly drops when the proxy offloads the connection to the switch (at sec. 30). In such test, the proxy splits in half an RTT of 100ms. Thus, when the proxy is in place, the end-points measure a 50ms RTT. Removing the proxy increases the end-points' measured RTT to 100ms, causing TCP to believe that one or more packets were lost. Therefore, TCP reduces the congestion window and triggers retransmissions, as shown by the reduced measured goodput.

To solve this issue, in preparation of the offload, our proxy implementation artificially increases the delay of the two connections that are going to be joined in one. The delay is gradually introduced in steps of 10ms, to give the TCP congestion control enough time to adapt its timeout values\footnote{The value has been empirically evaluated with experimental testing, using different TCP congestion control algorithm~\cite{siracusano2017tcp}, including the recent BBR implementation~\cite{cardwell2016bbr}.}.

\subsection{Internet deployment}

\begin{figure}
    \vspace{-1ex}

    \includegraphics[width=\columnwidth]{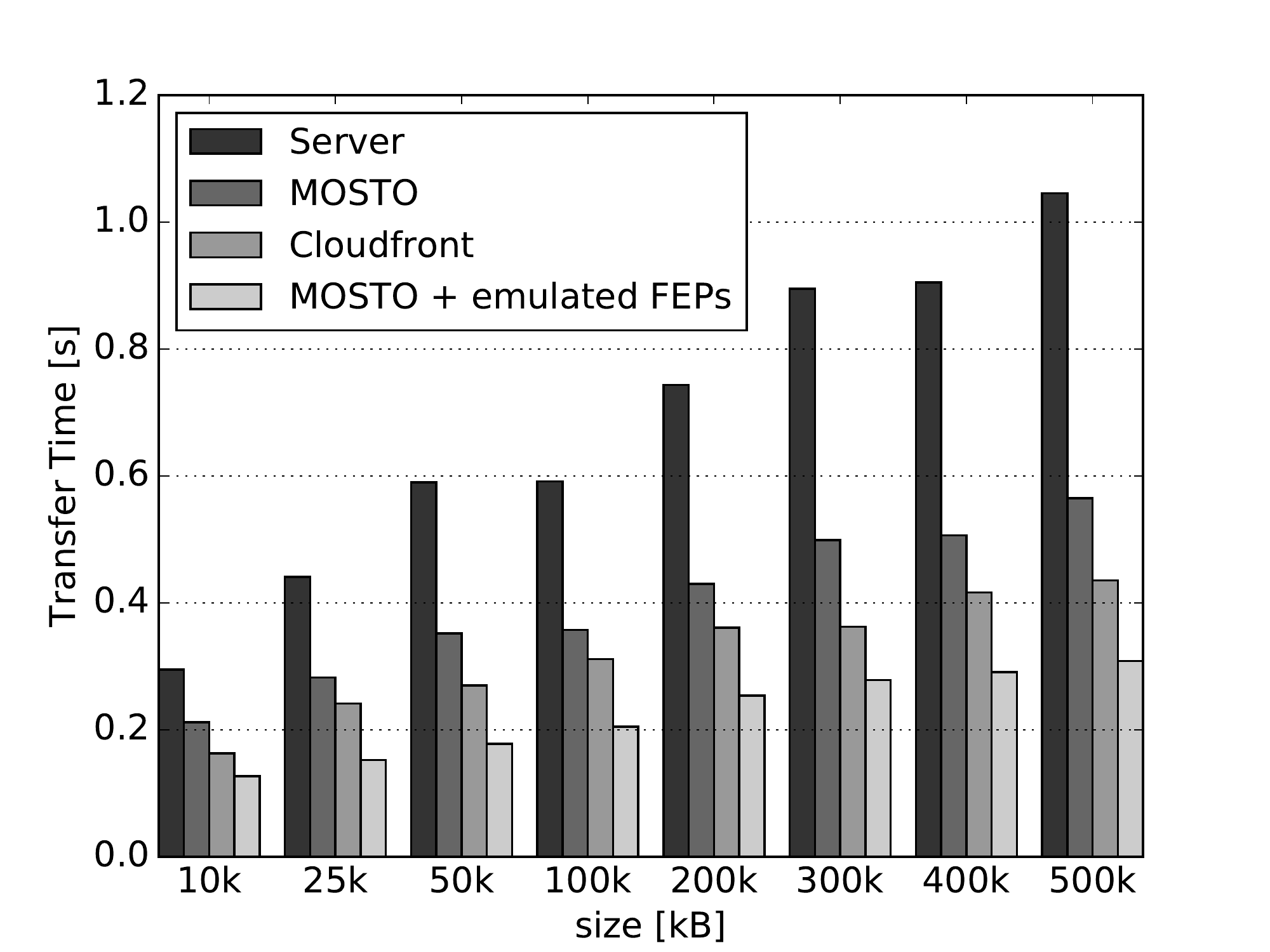}%
    \caption{MOSTO performance compared to client-server and Amazon CloudFront deployments}%
    \label{fig:ttc}%
        \vspace{-2ex}

\end{figure}

To validate the MOSTO implementation, we performed a proof-of-concept deployment using the Amazon EC2 cloud infrastructure. Since we use current cloud datacenters, we cannot take advantage of the optimizations described in Sec.~\ref{sec:issues}. Nonetheless, our measurements show the potential of the solution and the possible benefits it may provide.

\vspace{0.1em}
\noindent
\textbf{Transfer time}
We deploy a web client and a web server at EC2's datacenters in Ireland and California, respectively. Then, we measure the performance of a single client's HTTP request in different cases. First, we measure the direct connection performance. Second, we measure the same performance when using MOSTO to establish the connection. Finally, we subscribe to Amazon's CloudFront service, to provide access to the web server through the Amazon managed CDN service. 
CloudFront places front-end servers close to the clients. In our case, the front-end is actually co-located with the client, in the Ireland datacenter. In the case of a dynamic content request, a new TCP connection to the front-end triggers the opening of another TCP connection from the front-end to the web server, over which the request is relayed. Such connection is maintained open to serve other requests towards the same server. If no further requests are received, it is closed after 5s of inactivity. Thus, we evaluate also the performance of content delivery when using CloudFront in its operating best case, i.e., when there is a pre-established TCP connection with the web server.
Fig.~\ref{fig:ttc} summarizes the results for transfer sizes varying between 10 and 500 KBs. 

Our test shows that the direct connection is the slowest in completing the transfer, while CloudFront with a pre-established connection is clearly the fastest. MOSTO provides a transfer time which is 60-150\% better than a direct connection, but 10-35\% worse than CloudFront, depending on the packet size. Here, we point out that Fig.~\ref{fig:ttc} reports an average of the measured transfer times over 100 runs of the experiment. While the other measurements showed consistent results during the 100 runs, CloudFront measurements showed instead a much higher performance variation, despite the very convenient deployment location of the front-end.  For instance, in the case of a 500KBs transfer, only 5\% of the cases experienced the expected transfer time of one single RTT, i.e., about 180ms\footnote{Since the TCP connection is pre-established, its congestion window is already fully open and 500KBs can be transferred in a single round.}, while the median is instead 300ms, the third quartile rises to 600ms and the 95th percentile goes up further to 700ms.

We are unable to profile CloudFront performance; our hypothesis is that the measured performance variation may depend on effects such as head-of-line (HOL) blocking over the shared connection~\cite{qian2015tm}, the need to run the pre-established connection over a large distance, i.e., Ireland to California, and the fact that the connection is served by a complex (web) proxy instance shared with other customers.

MOSTO is largely unaffected by the above phenomena. First, the client requests use dedicated connections, avoiding the HOL problem. Second, using a proxy to split the end-to-end RTT mitigates the problems of a long delay TCP connection. Finally, even if we run a VM over a shared infrastructure, the VM has dedicated CPU resources to handle the client request. 

Furthermore, one should take into account the relatively small amount of locations available in EC2 to deploy proxies. In fact, the MOSTO controller has one obvious choice to place a proxy in between the client's and server's locations. In our tests, a single proxy instance is deployed in the EC2's Virginia datacenter. 
To verify that a larger number of FEPs on the path could improve MOSTO performance to a level comparable to CloudFront, we recreated the test scenario in a controlled testbed. In this last test, we deploy two additional proxies on the path between client and server, splitting the end-to-end delay in three components. The last bar of Fig.~\ref{fig:ttc} shows that in such case MOSTO performance is on average better, and in general comparable to that of an optimal CloudFront placement.

\vspace{0.1em}
\noindent
\textbf{Deployment cost} While it is in general very hard to correctly estimate the cost of running MOSTO, given the absence of business models for the envisioned Internet infrastructure, we decided anyway to perform a rough cost estimation using current cloud services prices. In particular, we take as reference the publicly available pricing for CloudFront and Amazon EC2. CloudFront currently charges in the order of $0,0075\$$ for 10k HTTP requests (independently from their size). Doing a conservative assumption that a request is large 500KBs\footnote{Very few HTTP requests have a size bigger than 60KB~\cite{overclockingYahoo}.}, we can express the cost in terms of dollars per terabit ($\$/Tb$). Therefore, a CloudFront deployment costs $0.1875 \$/Tb$. Runnig a VM that supports the performance required by a MOSTO proxy costs $0.108 \$/h$, i.e., that is roughly the current cost of an Amazon EC2 m4.large instance with two CPU cores and enhanced networking. Given that a MOSTO proxy can forward at least 10Gbps of traffic on a single core, an m4.large instance should be able to offer at least such performance, therefore costing $0.003 \$/Tb$.
From the test of Fig.~\ref{fig:ttc} we see that MOSTO performance is comparable or better to that of CloudFront when running 3 proxies, therefore, we consider the cost of such deployment to be $0.003\cdot3=0.009 \$/Tb$. While our goal is not to give an exhaustive cost comparison, this trivial calculation shows that a MOSTO deployment cost is little, and orders of magnitude smaller than current CDN solutions\footnote{It should be noted that we did not take into account other costs, such as data transfers.}.

\section{Discussion}
\label{sec:discussion}
The presentation of MOSTO done so far provides an understanding of the core mechanisms of the solution. Many aspects did not find space in this paper, therefore this section will briefly comment on them.

\vspace{0.1em}
\noindent\textbf{Reliability} The critical component for the reliability of MOSTO is the centralized controller. However, such controller is not on the data path of the system, meaning that a failure of the controller would not impact MOSTO's proxies operations. At most, if the failure lasts longer than the paths update period, proxies may start using not optimal paths.

\vspace{0.1em}
\noindent\textbf{Security} Using transparent interception of network traffic at the FEPs may introduce security concerns, since one could capture traffic of a third-party. We would therefore expect security mechanisms to avoid such a problem. For instance, a FEP's service could be allowed to intercept traffic only when the source/destination is a service's owned end-point~\cite{Stoenescu2015EuroSys}.

\vspace{0.1em}
\noindent\textbf{Mobility} In the current version of MOSTO we did not address user mobility. Nonetheless, we believe that mobility would have a minimal impact on the overall system performance. The motivation is twofold. First, MOSTO targets short flows, which having a short lifetime are unlikely to be affected by user mobility. Second, the impact of mobility depends on the closeness of the InP to the end user. Here, notice that our placement algorithm tends to put InPs in FEPs that are tipically located at 10s of ms from the user. That is, typically MOSTO would use Central Offices or Mobile Telephone Switching Offices as edge FEPs. In such a case, we expect a low number of flows would actually need to be migrated to a different InP. In such cases, related work to handle mobility in virtualized infrastructures could be applied~\cite{fmc1, fmc2}.

\vspace{0.1em}
\noindent\textbf{Containers} We use unikernels to build lightweight proxies, relying on the enhancements to TCP implementation described in ~\cite{siracusano2016fly}. Using Containers would have been trickier, since they rely on the TCP stack provided by the host OS kernel. Furthermore, at the time of writing, containers have longer boot times~\cite{190607}.

\vspace{0.1em}
\noindent\textbf{Transfer size discovery} Our path selection algorithm requires to know in advance the size of the transfer to find the optimal path. A possible approach to infer such size is to assign communication end-points, e.g., domain names or IP addresses, in dependence of the size of the transfers they will support. This way, a service subscribed to MOSTO can indicate that a given IP address will provide a majority of flows of a given size. More sophisticated solutions are also possible, including reading the application layer header for a new flow or collecting and analyzing statistical data.

\vspace{0.1em}
\noindent\textbf{Bandwidth and loss rate} While MOSTO optimizes transfers taking in consideration only the RTT of a connection, parameters such as bandwidth and loss rate may also play an important role. Taking them into account would require to change the path computation algorithm in the controller and is one of the possible future work directions.

\vspace{0.1em}
\noindent\textbf{CDNs footprint} MOSTO allows a CDN to reduce its geographical footprint. Instead of building thousands of presence points across the planet, a future CDN could rely on few central locations, e.g., datacenters. Thus, a MOSTO-like technology could be used to connect such datacenters to the edge of the network, avoiding the need to pre-deploy a large infrastructure.

\vspace{0.1em}
\noindent\textbf{Do it yourself} While traditional CDN providers could employ MOSTO as technology, the absence of pre-deployed infrastructures makes MOSTO a perfect technology for a more general use. For instance, MOSTO could be deployed by content providers to replace a CDN, but also by enterprises to connect geographically distant branch offices, or even by regular users to accelerate their connections towards any given destination.

\section{Related Work}
\label{sec:related}
Using proxies for TCP acceleration was originally proposed in \cite{appLevelRelays},  where a P2P overlay was used for the deployment. 
% In such an overlay each proxy may require network packets to have large detours. Furthermore, P2P nodes may be difficult to control and bring in the overlay a node churn issue. A different approach is proposed by 
Instead, \cite{ladiwala2009transparent} suggests the enhancement of routers to provide TCP proxy capabilities. 
% However, \cite{appLevelRelays} \cite{ladiwala2009transparent} assumes that routers opportunistically select connections that should be proxied among those that traverse the router. A route change would therefore break the connection. Furthermore, there is no mechanism that allows the explicit selection of which TCP connection to accelerate. Finally, the deployment of the system assumes the possibility to modify the routers traversed on the end-to-end path.
MOSTO applies the same approach but tackles scalability and deployability issues. Also, it shows how to leverage the future Internet infrastructures and related technological benefits.
% While the technique is well known and has been suggested in the past~\cite{ladiwala2009transparent}, deployability and scalability issues hindered its applicability. 

More generally, in terms of TCP optimization, past work has improved TCP performance for small flows by increasing the initial congestion window of TCP~\cite{allman2002increasing, chu2013increasing} and, for larger flows, proposing new algorithms for the congestion control during the TCP's steady state~\cite{ha2008cubic, cardwell2016bbr}. The Slow-Start optimization was addressed by proposals such as Quick-Start, XCP, RCP. However, these proposals are difficult to deploy since all routers on path have to be support them. \cite{briscoe2014reducing} presents a good survey of these mechanisms.

In terms of CDNs, \cite{Sundaresan:2013:CCA:2504730.2504741} highlights the dependence of web performance on the availability of CDN sites in a given geographical area. Which is the reason why very large infrastructures are being deployed~\cite{googleInfrExp}. MOSTO helps in reducing this dependence, as long as FEPs are available in the area. Notice that the approach of using third-party infrastructures, although specifically limited to the CDN service, is also discussed in \cite{mukerjee2016impact} and was envisioned in \cite{Ager:2011:WCC:2068816.2068870}. 
A related approach is used in VIA~\cite{via}: cloud datacenters are used to deploy Skype proxies to de-tour connections from parts of the network that are experiencing performance issues.

Specific mechanisms addressed in this paper are also covered in some related works. End-user mapping is discussed in \cite{chen2015end} and \cite{Stoenescu2015EuroSys}. Jitsu~\cite{jitsu} presents a reactive booting mechanism for VMs hosting TCP servers. ClickOS \cite{martins2014clickos} shows several network functions implemented as unikernels and Miniproxy~\cite{siracusano2016fly} specifically implements a TCP proxy unikernel.

\section{Conclusion}
\label{sec:conclusion}
We presented the design and implementation of MOSTO, a CDN service for the acceleration of dynamic web content delivery. During the design of MOSTO we considered a number of opportunities enabled by the deployment of SDN and NFV technologies in the Internet, and showed how the design of MOSTO could benefit from them. In this process, we identified as relevant functions: the reactive service instantiation; network measurement as a service; and programmable switches' forwarding entries. Such functions, as well as the technologies that are required to leverage them, can be easily integrated in the systems design, as demonstrated by our proof-of-concept implementation.

\section*{Acknowledgment}
We thank Mathias Niepert for his help in motivating this work in its early phases.
This paper has received funding from the European Union's Horizon 2020 research and innovation programme under grant agreements No. 671566 ("Superfluidity") and No. 671648 ("VirtuWind").
This paper reflects only the authors' views and the European Commission is not responsible for any use that may be made of the information it contains.

\small{
\bibliographystyle{abbrv}
%\bibliography{biblio}
\bibliography{bib_short.bib}
}

% {%\scriptsize

% \printbibliography
% }

\section{Appendix A: path computation}

To explain the algorithm, we first introduce some more notation. Let $n$ be the number of proxies, denoted $p_1,\ldots,p_n$. Let $m = n \cdot (n-1)$ be the number of proxy pairs or \emph{links}, denoted as $\ell_1,\ldots, \ell_m$ and sorted by nondecreasing distance $d(\ell_h)$. We denote with $G_h$ the proxy network where all links $\ell_{h+1},\ldots,\ell_{m}$ have been removed. Let
$
	D(i,j,h)
$
be the shortest length of all paths connecting $p_i$ and $p_j$ in $G_h$. If there is no such path, then $D(i,j,h) = \infty$. The dynamic programming algorithm computes $D(i,j,h)$ for all values of $i,j,h$, using the recursion
\begin{align*}
	D(i,j,h) = \min\{ & D(i,j,h-1) ,\\
	{} &  D(i,a_h, h-1)+d(\ell_h) + D(b_h,j,h-1) , \\
	{} & D(i,b_h, h-1)+d(\ell_h) + D(a_h,j,h-1) \} \ ,
\end{align*}
where $\{a_h,b_h\} = \ell_h$ are the two nodes connected by link $\ell_h$. This recursion formalizes the observation that the shortest path from $p_i$ to $p_j$ in $G_h$ either does not use link $\ell_h$, or it uses $\ell_h$ in one of the two possible directions. Together with the base cases $D(i,i,h) = 0$ for all $i$ and $D(i,j,0) = \infty$ for all $i,j$ the complete set of path lengths $\{D(i,j,h) \mid 1 \leq i,j \leq n, 1 \leq h \leq m \}$ and the corresponding paths can be computed. This represents the desired Pareto-front, because for every possible value of the maximum link distance the set contains the shortest path with this particular maximum distance.

The runtime for this computation is proportional to the number of combinations of $i,j,h$, which is $\Theta(n^2 m) = \Theta(n^4)$, because for each such combination we need to apply the recursive formula or a base case once. This runtime is needed both in the worst case and in the best case.

In what follows we describe a version of the dynamic programming algorithm which has the same worst case runtime but improves the best case as well as the typical practical runtime. 
For any link $\ell_h = \{a_h,b_h\}$ we define the set $A_h$ as the set of nodes $p$ whose shortest path to node $a_h$ in $G_h$ contains the link $\ell_h$. Conversely, we define the set $B_h$ to contain the nodes whose shortest path to $b_h$ in $G_h$ is via $\ell_h$. 

\begin{lemma}
	For any $h=1,\ldots,m$ it holds that $A_h \cap B_h = \emptyset$.
\end{lemma}

\noindent
\textbf{Proof.} If there is any node $q$ both in $A_h$  and $B_h$, then the shortest path from $q$ to $a_h$ in $G_h$ is via node $b_h$. The sub-path from $q$ to $b_h$ is the shortest possible one in $G_h$, thus by assumption containing $a_h$. Thus, there is a loop in the shortest path from $q$ to $a_h$, which is a contradiction because all distances are positive. \hfill{$\square$}

\begin{lemma}
	For any $i,j,h$, if the shortest path from $p_i$ to $p_j$ in $G_h$ traverses the link $\ell_h$ in the direction $(a_h,b_h)$, then $p_i \in B_h$ and $p_j \in A_h$. Symmetrically, if the path traverses $\ell_h$ in the direction $(b_h,a_h)$, then $p_i \in A_h$ and $p_j \in B_h$.
\end{lemma}

\noindent
\textbf{Proof.} If the shortest path from $p_i$ to $p_j$ in $G_h$ traverses $(a_h,b_h)$ in that direction, then it contains a sub-path from $p_i$ to $b_h$ via $\ell_h$ and a sub-path from $p_j$ to $a_h$ via $\ell_h$. As these paths are shortest in $G_h$, $p_i \in B_h$ and $p_j \in A_h$. The other sentence of the lemma follows by symmetry. \hfill{$\square$}

For each pair of nodes $p_i,p_j$ our algorithm maintains a list $L_{ij}$ of paths from $p_i$ to $p_j$. This list is initially empty, and at any time it is sorted by non decreasing maximum link distance appearing on the stored paths. The algorithm runs in $m$ iterations numbered $1,\ldots,m$, each time adding for each node pair $p_i,p_j$ the shortest connecting path in $G_h$ to the list $L_{ij}$, but only if this path is not already contained, that is, only if that path uses the link~$\ell_h$.

To do this efficiently, the algorithm in each iteration $h$ first computes the sets $A_h$ and $B_h$ using the above recursion. Then, for each node pair $p_i \in A_h$ and $p_j \in B_h$ it computes the shortest path from $p_i$ to $p_j$ in $G_h$, again using the above recursion. In case this path uses $\ell_h$ it is appended to the list $L_{ij}$. Other node pairs than those in $A_h \times B_h$ do not need to be considered due to the above lemma.

There are $m$ iterations, and constructing $A_h$ and $B_h$ in each iteration requires time~$\Theta(n)$. In the worst case $A_h$ and $B_h$ each contain exactly half of all nodes, and in that case there are $\frac{1}{4} n^2$ candidate node pairs to check. In the best case one of the sets is empty, and then there is no node pair to check. Thus, the worst cast runtime remains $O(n^4)$, while the best case is $\Omega(n^2)$. The runtime needed in practive is depending on the graph structure, but in experiments with various graphs we observed a vast improvement of the runtime as compared to the straightforward dynamic programming approach.

% \textit
% {
% Note: The data structure to represent the paths during the dynamic programming algo is also not completely trivial; if there is some space Tobias can make a description of it.
% }

\end{document}